# Optimization method to eliminate the influence of the conical acoustic lens on the transmission of laser beam using ZEMAX

Xianlin Song [a, #, *], Zouhua Chen [b, #], Aojie Zhao [a], Bo Li [a], Jinhong Zhang [a], Qiming He [a], Jianshuang Wei [c, d], Lingfang Song [e]

[a] School of Information Engineering, Nanchang University, Nanchang 330031, China;
[b] College of Sciences, Nanchang University, Nanchang 330031, China;
[c] Britton Chance Center for Biomedical Photonics, Wuhan National Laboratory for Optoelectronics-Huazhong University of Science and Technology, Wuhan 430074, China;
[d] Moe Key Laboratory of Biomedical Photonics of Ministry of Education, Department of Biomedical Engineering, Huazhong University of Science and Technology, Wuhan 430074, China;
[e] Nanchang Normal University, Nanchang 330031, China;
[#] equally contributed to this work

**ABSTRACT**

Photoacoustic tomography is a new medical imaging technology with the advantages of high resolution, high contrast and high penetration depth. There are three common photoacoustic imaging methods in practical applications: photoacoustic microscopic imaging (PAM), photoacoustic computed tomography (PACT), photoacoustic tomography (PAE). As an important branch of photoacoustic imaging, photoacoustic microimaging combines high contrast of optical imaging with high resolution of ultrasonic imaging. In photoacoustic microimaging system, acoustooptic coupling prism is a very important component, which is usually composed of irregular prism and spherical concave acoustic lens at the bottom. Its function is to carry out optical transmission and ultrasonic detection. The ultrasonic depth of field of spherical concave acoustic lens is limited. In order to overcome this defect, researchers propose to use conical concave acoustic lens to produce Bessel sound beam to realize large depth of field ultrasonic detection. But the conical concave acoustic lens affects laser focusing and imaging. In order to solve this problem, we propose an optimization method to eliminate the influence of conical concave acoustic lens on beam transmission. A calibration mirror is added to the acoustooptic coupling prism with conical concave acoustic lens at the bottom, and the deterioration of the cone concave acoustic lens to the beam transmission is eliminated by optimizing the surface shape and thickness of the calibration mirror by Zemax. The optimization effect is evaluated by analyzing the spot. The simulation results show that the optimization method can eliminate the influence of the conical concave acoustic lens on the beam transmission, make the focal point and the focal point keep the coaxial focus, and improve the detection efficiency of the photoacoustic signal. This work is of theoretical significance for the systematic study of large depth of field photoacoustic microimaging.

## 1. INTRODUCTION

As early as 1880, Bell discovered the phenomenon of photoacoustic conversion, and in a report to the American Association for the Advancement of Science called this physical phenomenon the "photoacoustic effect" [1]. In 2005, Maslov et al. [2] used photoacoustic microscopy imaging technology for blood vessel imaging for the first time and obtained clear images of blood vessel structure. The system uses a 50 MHz spherical focused ultrasound detector to detect the signal, and the laser uses a dark-field radiation mode to achieve the coincidence of the excitation light focus and the acoustic detection focus, achieving high horizontal and vertical resolution, and achieving acoustic resolution The photoacoustic microscopy imaging has opened up the rapid development of photoacoustic imaging technology[3][4]. The theoretical basis of the photoacoustic effect is the use of pulsed lasers to irradiate biological tissues, which will produce periodic temperature changes inside the tissues, which will then produce elastic thermal expansion. This transient elastic thermal expansion of tissues will generate pressure waves (ultrasound), thereby generating acoustic signals. Such acoustic signals are called photoacoustic signals. Different tissue components produce photoacoustic signals of different intensities due to the difference in light absorption, and the light absorption distribution image in the tissue is received and reconstructed by the ultrasound probe, that is, photoacoustic imaging. There are many ways to realize photoacoustic imaging technology, among which two important ways are tomography and microscopic imaging. Tomography uses diffused light to excite biological tissues, the imaging depth can reach several centimeters, and the spatial resolution can reach tens of microns [5]. Microscopic imaging uses focused lasers to excite biological tissues or uses focused ultrasound sensors to detect ultrasound signals in a specific area, and the lateral resolution can reach micrometers or even submicrometers [5].

## 2. MODEL AND METHOD

The optical design software Zemax is used to realize the optimal design of the collimating mirror. The result of the optimal design is determined by the system's initial structure, variables and optimized operands. By changing variables and optimizing operands and their weights, the results of the optimized design meet the requirements.

### 2.1 Steps to optimize design

The process of optimizing design can be divided into the following steps:

(1) Establish the initial lens model according to the initial parameters: set the entrance pupil diameter to 10 mm, the field of view to 0°, the wavelength to 0.633 μm.

(2) Before optimizing the design of the calibration mirror, set the surface type of the front surface of the calibration mirror to Odd Asphere and the radius of curvature to infinity, at the same time set the surface type of the back surface of the acoustic lens to Odd Asphere and the radius of curvature to infinity. The material of the calibration mirror is set to SILICA. After establishing the initial model, set the variables: set the thickness of surface 3 and the thickness of the first Odd Asphere as variables, then set Par 1 of the first Odd Asphere and Par 1 of the second Odd Asphere as variables;

(3) Optimization of geometric aberration and transfer function: First, optimize the design of the astigmatism of geometric aberration. After the geometric aberration is optimized, if the imaging quality of the off-axis field of view still fails to meet the requirements, the transfer function must be optimized again. When optimizing a system with aberrations greater than 2 to 5 wavelengths, the geometric transfer function (GMMTT, GMTS, GMTA) is selected as the operand for image quality evaluation. What's more, the image quality can be further improved through physical communication (MTFT, MTFS, MTFA).

### 2.2 Establishment of initial lens model

Set the initial structure parameters of the lens in Zemax: including entrance pupil diameter, field of view, wavelength, and lens material. Then in the lens data editor, set the surface type of the front surface of the calibration mirror to Odd Asphere and the radius of curvature to infinity, at the same time set the surface type of the back surface of the acoustic lens to Odd Asphere and the radius of curvature to infinity. Setting the lens material to SILICA in the column of the first Odd Asphere data. The thickness of the surface 3 and the thickness of the first Odd Asphere are set as variables. Finally, set the Par 1 of the first Odd Asphere and the second Odd Asphere as variables. Then open the default evaluation function and set the optimization objective of the evaluation function to the spot radius, and the pupil sampling is Gaussian Quadrature. Finally, select local optimization in Zemax and check the automatic update option.

Optimizing the astigmatism of the system to be smaller first, then judge the main aberrations of the current system by combining the point diagram and the MTF curve graph, and the corresponding methods are adopted to correct the main aberrations that currently exist, and then optimized again. For example, the optimized design can be completed by adding optimized operands and their weights.

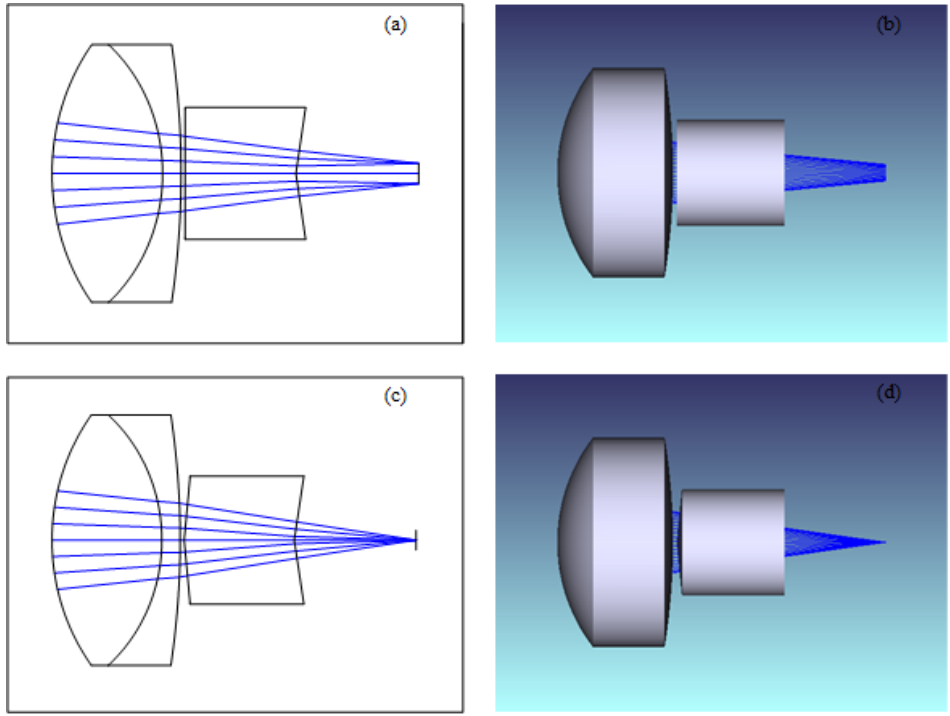

Figure 1. (a) is the 2D view of the lens system before the calibration mirror is optimized; (b) is the shadow model before the calibration mirror is optimized; (c) is the 2D view of the lens system after the calibration mirror is optimized; (d) is the shadow model after the calibration mirror is optimized.

## 3. RESULTS AND DISCUSSION

Table 1. Data of the optimized lens system.

| Surface | Surface Type | Radius | Thickness | Glass | Semi-Diameter | Par 1 |
|---|---|---|---|---|---|---|
| OBJ | Standard | Infinity | Infinity | | 0.000 | |
| 1 | Standard | 20.900 | 12.000 | N-BAF10 | 12.700 | |
| 2 | Standard | -16.700 | 2.000 | N-SF6HT | 12.700 | |
| 3 | Standard | -79.800 | 0.440 | | 12.700 | |
| 4 | Odd Asphere | Infinity | 12.020 | SILICA | 6.350 | 0.100 |
| 5 | Odd Asphere | Infinity | 13.320 | | 6.500 | 0.162 |
| IMA | Standard | Infinity | | | 1.000 | |

It can be seen from Table 1 that the first Odd Asphere after the calibration mirror is optimized it's Par 1 get to 0.1. The second Odd Asphere it's Par 1 has a magnitude of 0.162. The thickness of the first Odd Asphere is the number of 12.020, and the thickness of the second Odd Asphere is the number of 13.320.

After many rays of light emitted from one point are refracted by the optical system, due to the existence of aberrations, the intersection point with the image plane no longer converges at one point, but forms a diffuse spot scattered in a certain range, which is called spot diagram. One of the methods for judging the image quality of the system is to observe the density of spot in the spot diagram.

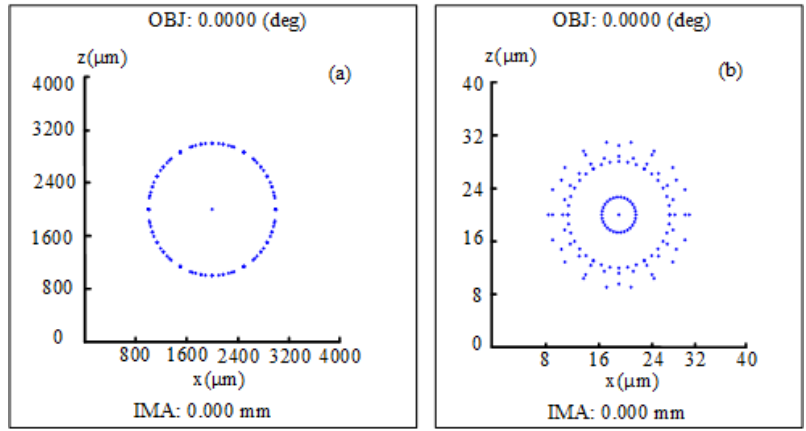

Figure 2. (a) is spot diagram of the lens system before optimization under 0° field of view value; (b) is spot diagram of the lens system after optimization under 0° field of view value.

It can be seen from Figure 2. (a) that the spherical aberration of the system before the optimization of the calibration mirror is very large, the root mean square radius of the spot under the 0 °field of view is 993.842 μm, and the geometric radius is 1000.98 μm. From Figure 2. (b) it can be seen that after the calibration lens is optimized, the spot is concentrated, the spherical aberration is greatly improved, and the root mean square radius of the spot get to 8.091 μm under the 0 °field of view, and the geometric radius of the spot get to 11.087 μm. It can be seen that after the calibration mirror is optimized, the spherical aberration of the system is significantly improved, and the overall imaging quality of the system is greatly improved.

The optical imaging system uses the optical transfer function to evaluate its imaging quality. The MTF value is the modulation transfer function that indicates the transmission of various frequencies, and comprehensively evaluates the sharpness, contrast and resolution of the lens. For general lenses, when MTF>0.3, it is considered to be clearly recognizable, when MTF>0.6 the image is considered good, and when MTF>0.8 the image quality is considered very good.

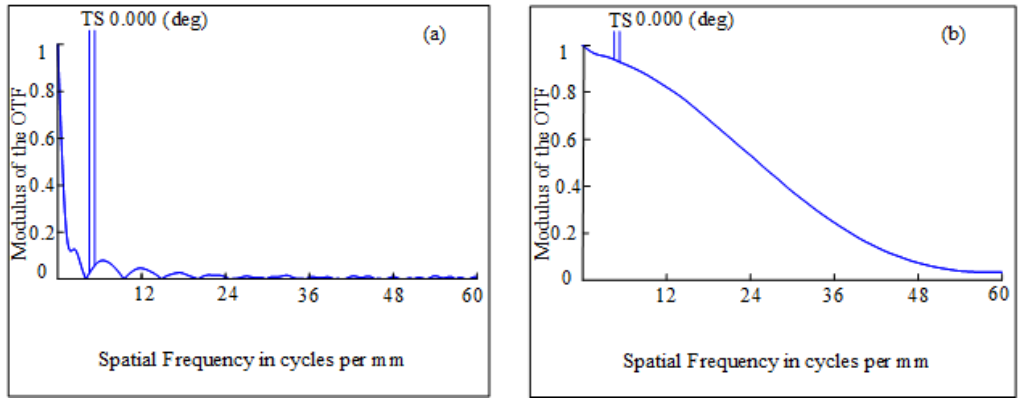

Figure 3. (a) is the MTF of the meridian and sagittal direction before optimization under 0° field of view value; (b) is the MTF of the meridian and sagittal direction after optimization under 0° field of view value.

## 4. CONCLUSION

In this paper, by using the optical design software Zemax, on the basis of geometric optics and primary aberration theory, the optimal design method of collimating mirror is discussed and proposed, which eliminates the influence of conical concave acoustic lens on beam transmission. The lens system imaging before and after the optimization of the calibration lens is simulated by numerical simulation: from the simulation results, after the calibration mirror is optimized, the spherical aberration of the system is greatly reduced,. The root mean square radius of the spot under the 0 °field of view changes from 993.842 μm to 8.091 μm, and the geometric radius changes from 1000.98 μm to 11.087 μm; the MTF curve is obviously improved, the cut-off frequency is increased by nearly 15 times, and the MTF value of the meridian direction and sagittal direction under the 0 ° field of view are between 0.9 and 1; The proposed optimization method of the collimating mirror has important theoretical guiding significance for the study of the large depth of field photoacoustic microscopy imaging system.

## REFERENCES

[1] Bell, A. G., "Upon the production and reproduction of sound by light," American Journal of Science, 20(118), 305-324 (1880).
[2] Maslov, K., Stoica, G. and Wang, L. V., "In vivo darkfield reflection-mode photoacoustic microscopy," Optics Letters, 30(6), 625-627 (2005).
[3] Zhang, H. F ., Maslove, K., Sivaramakrishnan, M., Stoica, G. and Wang, L. V., "Imaging of hemoglobin Oxygen saturation variations in single vessels in vivo using photoacousticmicroscopy," Applied Physics Letters, 90(5), 053-901 (2007).
[4] Maslov, K., Zhang, H. F., Hu, S. and Wang, L. V., "Optica resolution photoacoustic microscopy for in vivo imaging of single capillaries," Optics letters, 33(0), 020-021 (2008).
[5] Wang, L. V. and Hu, S., "Photoacoustic tomography：in vivo imaging from organelles to organs," Science, 335(6075), 1458-1462 (2012).